Title:

# HIV DRUG RESISTANCE: PROBLEMS AND PERSPECTIVES

Author: Pleuni S Pennings

Affiliation: Stanford University

Running title: HIV DRUG RESISTANCE REVIEW

Submitted to: Infectious Disease Reports
    http://www.pagepress.org/journals/index.php/idr/index

Acknowledgements: I would like to thank Daniel Rosenbloom, Alison Hill, Stefany Moreno, Jonathan Li and Nandita Garud for useful comments on an earlier version of this manuscript.

Version: Jan 20th 2013

Corresponding author:
Pleuni Pennings,
email: pleuni@stanford.edu OR pleuni@dds.nl
address:
Department of Biology
371 Serra St.
Stanford University
Stanford, CA 94305-5020

Keywords: HIV, drug resistance, resistance, antiviral therapy, ART

# HIV DRUG RESISTANCE: PROBLEMS AND PERSPECTIVES

## Abstract

Access to combination antiretroviral treatment (ART) has improved greatly over recent years. At the end of 2011, more than eight million HIV infected people were receiving antiretroviral therapy in low-income and middle-income countries. ART generally works well in keeping the virus suppressed and the patient healthy. However, treatment only works as long as the virus is not resistant against the drugs used. In the last decades, HIV treatments have become better and better at slowing down the evolution of drug resistance, so that some patients are treated for many years without having any resistance problems. However, for some patients, especially in low-income countries, drug resistance is still a serious threat to their health. This essay will review what is known about transmitted and acquired drug resistance, multi-class drug resistance, resistance to newer drugs, resistance due to treatment for the prevention of mother-to-child transmission, the role of minority variants (low-frequency drug-resistance mutations), and resistance due to pre-exposure prophylaxis.

## Introduction

More and more HIV patients have access to combination antiretroviral treatment (ART). ART generally works well in keeping the virus suppressed and the patient healthy. However, treatment only works as long as the virus is not resistant against the drugs used. When the first antiretrovirals came on the market in the 1980s, drug resistance was a certain outcome for all patients, and the duration of successful treatment was limited. Nowadays, some patients are treated for many years without having any resistance problems, while for others, drug resistance is a serious threat to their health.

Throughout this essay I will contrast the situation in high-income countries, where combination therapy has been common since the late 1990s (North America, Europe, Japan, Australia), with the situation in low- and middle-income countries, where, in some areas, treatment has only recently become available. At the end of 2011, more than eight million people were receiving antiretroviral therapy in low- and middle-income countries. The number of people on treatment increased most rapidly in sub-saharan Africa, from just 100,000 patients on treatment in 2003 to 6.2 million in 2011 (UNAIDS, 2012). These numbers are impressive and have changed the lives of HIV patients in those regions. However, the situation of patients in low-income countries is still different in several ways from the situation of patients in high-income countries. For example, viral load monitoring and viral genotyping, which are standard practice in high-income countries, are almost completely unavailable in low-income countries.

In order to cover different aspects of drug resistance, this essay is split in seven parts. (1) I will start by describing the problem of **transmitted drug resistance.** Next (2), I will elaborate on **acquired drug resistance** during ART, which is more common than transmitted drug resistance. I will then (3) describe which patients are at risk of developing **multi-class drug resistance** (MDR) and discuss briefly the treatment options that are available to them. (4) In the last decade, several new drugs, in existing as well as new drug classes, have become available. I will describe what is known about **resistance to some of these new drugs**. (5) In low- and middle-income countries, many patients are exposed to antiretroviral drugs in the context of **prevention of mother-to-child-transmission** (PMTCT). I will explain how PMTCT can lead to a high risk of drug resistance. (6) Next, I will look at what is known about drug-resistance mutations which are present at low frequency in the patient, which are known as **minority variants**. (7) Finally, I will touch upon the issue of drug resistance due to the use of antiretroviral drugs to prevent HIV infections (**pre-

**exposure prophylaxis** or PrEP).

## 1. Transmitted Drug Resistance

Drug-resistant HIV strains can be transmitted from one patient to another. Due to such transmitted drug resistance, a newly infected patient may carry a drug-resistant virus even though he or she has not yet used antiretroviral drugs. From the early days of HIV treatment, researchers have feared that drug-resistant strains would reach high frequencies among newly infected patients, rendering certain drugs entirely useless. Fortunately, this has never happened with HIV drugs. In comparison, many malaria drugs have been withdrawn from use by national authorities because of widespread transmission of drug resistant malaria parasites (Read and Huijben, 2009). Transmitted drug resistance does occur in HIV, but the numbers have remained relatively low. In high-income countries, between 7 and 17% of newly infected patients carry at least one major drug-resistance mutation, usually a mutation that confers resistance to one of two drug classes: nucleoside reverse transcriptase inhibitors (NRTI) or non-nucleoside reverse transcriptase inhibitors (NNRTI) (WHO, 2012, Frentz et al., 2012). The transmission of drug resistance to the third major drug-class (protease inhibitors, PIs) is less common.

In middle and low-income countries, the levels of transmitted drug resistance are lower than in high-income countries, but this will likely change with time. In 2006, the member states of the United Nations agreed to aim for universal access to treatment in 2010 (UNAIDS, 2007). The rapid increase of availability of ART in middle- and low-income countries following this decision is referred to as the "rollout of ART". There is evidence for a rapid increase in transmitted drug resistance in the years after rollout of ART in east and southern Africa (Gupta et al., 2012). This is not surprising, because ART went from being virtually non-existent to being quite common in these regions. The prevalence of transmitted drug resistance in middle- and low-income countries is estimated to be around 7% (WHO, 2012).

Even though transmitted drug resistance is somewhat less common in low-income countries than in high-income countries, the impact may be higher for patients in low-income countries. In high-income countries, it is standard practice to genotype the virus before starting treatment, to determine whether resistance mutations are present. If such genotyping is done and a fully-active combination of drugs is chosen, treatment success of patients with transmitted drug resistance is very good. For example, Wittkop et al. (2011) estimated that among European patients with transmitted drug resistance who were treated with a fully active combination of drugs, 95% had fully suppressed viral load after one year. Of the patients who were treated with an insufficiently strong regimen, 85% had fully suppressed viral load after one year.

In low-income countries, viral genotyping is usually not available. Therefore, patients with transmitted drug resistance may start insufficiently strong ART regimens, which will be less effective in reducing the viral load, which, in turn, can lead to the evolution of multi-class drug resistance. In addition, even if testing is done, fewer second-line treatment options are available for patients in low-income countries.

## 2. Acquired Drug Resistance during ART

Since 1996, standard treatment of HIV has been with a combination of three drugs. In principle, many combinations can be created from the long list of antiretroviral drugs that are on the market. In practice, only a few combinations are used for most patients and recommended as first-line treatment for treatment-naive patients. The most common combination therapy consists of an NNRTI and two NRTIs, which is available as a one-pill-a-day treatment. Also common in high-income countries is the combination of a protease inhibitor (PI) with two NRTIs. This combination is only used as second-line therapy (when standard drugs fail) in low-income countries and not available as a one-pill-a-day regimen. In order to make PI-based treatment as effective as NNRTI-based treatment, the blood levels of the PI need to be increased by adding a small dose of another protease inhibitor (ritonavir), which is called boosting. Nowadays, boosted PI regimens are more common than unboosted PI regimens. Recently, a third option has been added to the list of recommended combinations for treatment-naive individuals: two NRTIs combined with an integrase strand transfer inhibitor (INSTI). This combination is also available as a one-pill-a-day regimen. The evolution of drug resistance on these combination treatments is much less likely than the evolution of drug resistance on treatments consisting of just one or two drugs, as were common in the late 1980s and the first half of the 1990s. Still, drug resistance can evolve during treatment.

The clearest pattern of acquired drug resistance in patients on ART is that the percentage of patients with drug resistance goes up steadily with time on treatment. For example, in a meta-analysis of studies from resource-limited settings, Stadeli and Richman (2012) find that 7% of patients who have been on ART for 6-11 months have resistance, 11 % after 12 -23 months and 21% after 36 months or more. A similar effect has been reported for patients in high-income countries. For example, a large study from the UK (Philips, 2010) reports that the percentage of patients with at least one drug-resistance mutation increases from 11 to 14 to 18% after four, six and eight years respectively for patients on NNRTI-based treatment. This study shows that even a patient whose viral population did not evolve resistance during six years of treatment has a probability of around 2% to acquire resistance during the next year of treatment.

In general, treatments based on NNRTIs or unboosted PIs are more susceptible to resistance than treatment based on a ritonavir-boosted PI (bPI) (Kempf et al., 2004). bPI regimens are less susceptible to resistance, partly because resistance to the bPI itself is unlikely to evolve, but also because, in the presence of the bPI, it is unlikely resistance evolves to the other drugs that the patient is taking (NRTIs). This protective effect of the bPI can be explained, in part, by the low viral loads achieved on bPIs. However, Kempf et al. (2004) showed that NRTI resistance is unlikely to evolve even in patients with viremia (unsuppressed virus) if they are taking a bPI. The mechanism behind this observation is not yet understood.

NNRTI-based ART is more common in low-income countries than in high-income countries, which means that there is a relatively high risk of drug resistance in low-income countries. In addition, if a patient's virus acquires drug resistance on NNRTI-based ART, it is less problematic in high-income countries because it will be discovered faster due to regular viral load monitoring and viral genotyping, so that the patient can be switched to second-line treatment when necessary.

There is a strong association between acquired drug resistance and sub-optimal adherence to the treatment regimen. For example, Lima et al. (2008) show that drug resistance is more than twice as common in patients with 80-95% adherence when compared to patients with adherence levels of 95% or higher. It is clearly documented that structured treatment interruptions lead to drug resistance (e.g., Danel et al., 2009), likely because an interruption allows for growth of the viral population and leads to a higher abundance of resistance mutations. When treatment is started again, these resistance mutations can quickly rise in frequency and lead to failure (Pennings, 2012). It seems plausible that non-adherence leads to drug resistance in exactly same way. In addition, sub-optimal adherence can lead to periods of effective monotherapy (the presence of just one drug above the minimally effective concentration) when drugs have very different half-lives. Effective monotherapy is most likely to occur in patients on NNRTI-based treatments, because NNRTIs typically have longer half-lives than NRTIs (Bangsberg 2006). However, the importance of effective monotherapy for resistance is debated (Pennings 2012).

In low-income countries, there is concern about resistance due to unplanned treatment interruptions when patients are faced with stock-outs at the hospital or pharmacy. For example, Marcellin et al. (2008) show that treatment interruptions occur in Cameroon due to drug shortages in hospitals.

## 3. Multi-class drug resistance

Multi-class drug resistance typically occurs when a virus that is resistant to one drug acquires resistance to another drug. In principle, it is possible that a virus acquires multiple drug-resistance mutations at the same time, but data suggest that this is less common. For example, Harrigan et al (2005) analyze data from a large Canadian cohort and show that at any one time, there are more patients with drug resistance to one class of drugs than patients with resistance to more than one drug. On the other hand, data reviewed by Stadeli and Richman (2012) suggest that in resource-limited settings, multi-class drug resistance is more common and around three-quarters of patients with at least some resistance have multi-class drug resistance. This difference may be due to lack of monitoring in resource-limited settings (Gupta et al., 2009). Unfortunately, data on multi-class drug resistance and the effect of monitoring are scarce, but the general belief is that switching quickly after detection of the first drug-resistance mutation can prevent the accumulation of further drug-resistance mutations. If treatment is continued even though one of the drugs no longer works, the virus will likely evolve resistance to the other drugs.

The Plato II study (Nakagawa et al. 2012) showed that, in Europe, the prevalence of patients who had failed on all three major drug classes (NRTI, NNRTI and PI) increased steadily after 1996, but remained stable from 2005. This is probably because the incidence of multi-class resistance went down, which, in turn, can be attributed to improvements in monitoring, simpler and less toxic regimens, which led to better adherence, and better pharmacodynamics, which made regimes more robust to sub-optimal adherence (Lundgren, 2012). As the risk of multi-class drug resistance went down, the chances of successful treatment for patients with multi-class resistant virus improved over the last decade (Castagliola et al., 2012). This is mainly due to the fact that several new drugs entered the market, such as raltegravir (INSTI) and darunavir (PI). When a patient's virus has many resistance mutations and an uncommon combination of three or more drugs needs to be used, this is referred to as salvage therapy.

## 4. Resistance to the newer drugs

For a long time there were only three main drug classes available for HIV treatment: NRTIs, NNRTIs and PIs, and the drugs within these classes were characterized by extensive cross-resistance. If a patient had failed treatment on one NNRTI, it was unlikely that treatment with another would work. From 2003 multiple new drugs from old and new drug classes entered the market. The new drug classes are integrase strand transfer inhibitors (INSTI, such as elvitegravir and raltegravir), CCR5 antagonists (maraviroc) and fusion inhibitors (enfuvirtide).

There are large differences between the genetics of resistance to different drugs, and the new classes are no exception. For example, single mutations can confer resistance to the integrase inhibitors raltegravir and elvitegravir, whereas multiple mutations are needed to confer resistance to newer integrase inhibitors (DTG and MK-2048, Mesplède et al., 2012). A single mutation at position 143, 148 or 155 of the integrase gene can make the virus highly resistant against raltegravir. The resistance profile of elvitegravir is similar, and mutations at position 148 and 155 were also observed in patients who failed treatment with the "quad" pill, a one-pill-a-day regimen that contains combistat-boosted elvitegravir and two NRTIs (DeJesus et al., 2012).

The fusion inhibitor enfuvirtide has been available since 2003, but is not used as first-line therapy, partly because it has to be injected subcutaneously by the patient. Several mutations are known to confer resistance to enfuvirtide (Hirsch et al., 2008). Resistance against the CCR5 antagonist maraviroc comes in two distinct flavors. Either, the virus can accumulate mutations that allow it to use inhibitor-bound CCR5, or the virus can switch tropism and use CXCR4 instead of CCR5 as a co-receptor to enter the cell (Moore and Kuritzkes, 2009). The latter is more common because CXCR4-using variants can be present at relatively high frequencies even prior to treatment with a CCR5 inhibitor. A recent study based on deep-sequencing found CXCR4-using variants in more than 90% of patients, though at very low frequencies in many of them (Swenson et al., 2011).

## 5. Prevention of mother-to-child transmission (PMTCT)

Pregnant women in low-resource settings are often treated to prevent the transmission of HIV from the mother to her child. The simplest option, which is no longer recommended, is to use only nevirapine (NVP, an NNRTI). A single dose of nevirapine (sdNVP) reduces the probability that the child is infected perinatally, but leads to a high risk of drug resistance in the mother and in the child, if it becomes infected despite nevirapine. In a meta-analysis, Arrivé et al (2007), found that on average 36% of the women treated with sdNVP had detectable NVP resistance several weeks after the treatment and 53% of the children. Other studies have shown that women who are previously treated with sdNVP are more likely to fail therapy if they are later treated with NNRTI-based ART (Lockman et al 2010). For these women, it is better to start with bPI-based ART.

In Pennings (2012), I show that the high probability of resistance after sdNVP is likely due to the combination of monotherapy with a high abundance of pre-existing NNRTI mutations. The situation is somewhat better if women are treated with the NRTI zidovudine (ZDV) for at least a couple of weeks before receiving sdNVP (Arrivé et al., 2007). Treatment with ZDV reduces the viral load and therefore the abundance of NNRTI resistance mutations. Prior ZDV treatment therefore leads to a lower probability that NNRTI mutations increase in frequency due to sdNVP, (Pennings, 2012). The probability of resistance is also reduced if NVP monotherapy is avoided by adding a so-called "NRTI tail", i.e., a combination of two NRTIs (ZDV and 3TC) for one week after delivery (Arrivé et al 2007, Pennings, 2012). Alternatively, pregnant women may be treated with a three-drug combination therapy throughout their pregnancy and while they are breastfeeding, even if they are not eligible for treatment for their own health.

## 6. Minority variants

Standard genotyping assays can only detect variants that have a relatively high frequency in the patient (around 20% frequency or higher). In recent years, however, studies have started to use methods that can detect low-frequency variants (down to 0.1% frequency, depending on the assay) and the presence of such low-frequency resistance mutations is associated with an increased risk of virologic failure, i.e., the inability to achieve or maintain suppression of viral replication, due to resistance (Li et al., 2011). From an evolutionary perspective, this is entirely expected. If the mutation that confers resistance is already present in the patient before treatment starts, such mutation may increase in frequency rapidly after the start of treatment and lead to virologic failure. Nonetheless, studies on minority variants have also shown that treatment can be – and often is – successful in patients despite the presence of minority variants. This means that even if a mutation is present at high enough frequency to be detected, there is no guarantee that it will lead to virologic failure. One reason for this may be that even though the mutation leads to resistance to one drug, the other two drugs in the treatment can suppress the virus sufficiently. Another, yet untested, possibility is that the location of origin of the detected mutation (e.g., blood vs. spleen) determines whether a minority variant increases or decreases in frequency when treatment is started. More research is needed to understand under which circumstances minority variants lead to treatment failure. At the same time, it has become clear that treatment can fail in patients even if no minority variants are detectable, because new mutations can happen or emerge from the reservoir of latent cells (Li et al., 2011).

There is the hope that cut-off values for the frequency of known resistance mutations can be determined to guide treatment decisions (Gianella and Richman 2010). Such a cut-off value would mean that a mutation with higher abundance than this value indicates an increased risk of treatment failure, whereas the same mutation at an abundance below the cut-off doesn't. From an evolutionary perspective, it is improbable that a sharp cut-off value exists, since each resistant viral particle has an equal chance to cause treatment failure. The probability of treatment failure is therefore likely to grow roughly linearly with the abundance of a rare resistant variant. Of course, for clinical purposes, a cut-off value may still be determined depending on the probability of failure due to a resistance mutation at a given frequency and the benefits gained from prevented failures.

It may be possible to reduce the risk of failure due to minority variants, without knowing which patients carry them, by modifying the way treatment is started. For example, treatment could be started with a set of drugs that are not susceptible to drug-resistance (e.g., a bPI-based combination) and once the viral load is sufficiently reduced – and therefore the abundance of any minority variant is equally reduced – the patient can be switched to the treatment of choice, such as NNRTI-based treatment, which is cheaper and available as a co-formulated one-pill-a-day regimen. Such a modified start of treatment may be acceptable for all patients, and minority variant assays are not needed.

Assays for minority variants can be useful to choose a regimen for pre-treated patients. For example, the presence of low-abundance NNRTI mutations in women who were previously treated for PMTCT, predicted treatment failure when they started NNRTI-based ART (Boltz et al. 2011). Fisher et al. (2012) used a deep sequencing approach and detected minority variants in patients failing bPI-based ART. Swenson et al. (2011) used a deep sequencing approach to predict the success of treatment with a CCR5 inhibitor.

## 7. Drug resistance and PrEP

Pre-exposure prophylaxis (PrEP) is the use of antiretrovirals to prevent HIV infection. Trials have looked at the effectiveness of tenofovir (TDF) as a pill or a vaginal gel and Truvada (co-formulated tenofovir and emtricitabine, TDF/FTC) to prevent infections, with success in some, but not all trials. PrEP could, in principle, lead to increased levels of drug resistance in several ways. First of all, the prophylactic antiretrovirals may not work against TDF- or FTC-resistant HIV strains and could therefore allow infections with resistant strains, leading to a higher relative level of transmitted drug-resistance (Supervie et al. 2010). Secondly, if someone becomes infected but is still using PrEP, the antiretrovirals used for PrEP could select for resistance. However, in the early PrEP studies (Grant et al. 2010, Baeten et al. 2012, Abdool Karim et al. 2010), none of the infections that occurred during the trial were with drug-resistant strains, possibly because infections mainly occurred in participants who were not taking their drugs regularly.

Another risk occurs when patients start using PrEP while already infected with HIV. This has occurred several times in the early PrEP studies. In the iPrEx study (Grant et al. 2010) two patients on the TDF/FTC arm were already infected with HIV, but did not know about it. The result was that they were taking TDF/FTC while already infected, which led to resistance in at least one of the two patients, probably because the drugs are not strong enough to suppress established virus even though they can prevent infection. In the Partners PrEP study (Baeten et al. 2012) there were also two cases of resistance due to the use of PrEP by previously infected people. These results show that unrecognized infections can be a problem for those who take antiretrovirals for PrEP.

## Conclusions

HIV is known as a fast evolving virus. It earned this reputation in the 1980s when treatments were typically only effective for a few months, because drug resistance evolved so quickly. However, over the last decades HIV treatments have become better and better at slowing down the evolution of drug resistance. Although no patient is entirely protected from drug resistance, rates of acquired drug resistance are low, on the order of a few percent per year of treatment.

Clinical trials, combined with viral load monitoring and viral genotyping, have made it possible to find treatments that minimize the risk of acquired drug resistance. Even without a thorough understanding of the mechanisms of the evolution of drug resistance, it is straightforward to count the number of patients with viral failure and resistant virus in the various treatment-arms of a clinical trial, allowing for a steady improvement of treatment regimens from one trial to the next. On the other hand, finding the best way to prevent transmitted drug resistance is a much harder problem, mainly because the level of transmitted drug resistance is determined by many factors in a community, not just the individual patient. These factors are hard to capture in a trial, plus, such a trial would need to compare communities as opposed to patients, which is more costly. Fortunately, preventing acquired drug resistance has helped to keep levels of transmitted drug resistance relatively low.

We should not forget that the issue of HIV drug resistance is not yet solved. In fact, drug resistance is mostly a problem for patients who also have other problems. For example, a pregnant and HIV-positive woman who has no access to proper treatment, may get treated with sdNVP to prevent mother-to-child transmission. The use of sdNVP can lead to NNRTI resistance, which means that if regular treatment becomes available for her later, it may not even work. In addition, if she lives in an area where viral load monitoring is not available, she may stay on the failing treatment and acquire additional drug-resistance mutations, which, in turn, can compromise the usefulness of second-line treatments. To solve the problem of drug resistance, the availability of both drugs and monitoring needs to increase drastically.

## Funding

This work was supported by a long-term fellowship of the Human Frontier Science Program (LT000591/2010-L).

## List of abbreviations

ART: antiretroviral therapy, usually referring to three-drug regimen of two NRTIs and an NNRTI or a PI.
NRTI: nucleoside/nucleotide analogue reverse transcriptase inhibitor
NNRTI: non-nucleoside analogue reverse transcriptase inhibitor
PI: protease inhibitor
bPI: ritonavir-boosted protease inhibitor
INSTI: integrase strand transfer inhibitor
PMTCT: prevention of mother-to-child transmission
MDR: multi-class drug resistance
PrEP: pre-exposure prophylaxis
TDF: tenofovir (NRTI
FTC: emtricitabine (NRTI)
ZDV: zidovudine (NRTI)
3TC: lamivudine (NRTI)
NVP: nevirapine (NNRTI)
sdNVP: single-dose nevirapine